\documentclass[12pt]{iopart}
\usepackage{graphicx}
\begin{document}
\title{Creation, doubling, and splitting, of vortices  in intracavity second harmonic generation}

\author{O-K Lim\dag, B Boland\dag, M Saffman\dag and W. Krolikowski\ddag}
\address{\dag Department of Physics,
University of Wisconsin, 1150 University Avenue,  Madison, Wisconsin 53706, USA}
\address{\ddag
Laser Physics Centre, 
Research School of Physical Sciences and Engineering,
Australian National University,
Canberra ACT 0200, Australia.
}

\begin{abstract}
We demonstrate generation and frequency doubling of unit charge vortices in a linear astigmatic resonator. 
Topological instability of the double charge harmonic vortices leads to well separated vortex cores that are shown to rotate, and become anisotropic,  as the resonator is tuned across resonance.  
\end{abstract}
 \date{\today}
\maketitle

Optical vortices are topological objects whose transformation properties under propagation in linear and nonlinear optical media have been the subject of much recent work\cite{vortexbook}. The  vortex  
charge of a beam, defined as the closed loop contour integral of the wave phase modulo $2\pi,$ is generally a conserved quantity under linear propagation in free space. 
Optical vortices occur naturally in speckle fields\cite{zeldovich}, and can be generated in a controlled fashion using diffraction from holographic plates\cite{bazhenov,heckenberg}. They can also be generated in lasers\cite{rigrod,weisshelical} and cavities with nonlinear elements\cite{lippi}, while high order vortex modes have been observed in active cavities with field rotating elements\cite{phscripta,volostnikov}. 
Astigmatic optical elements as well as nonlinear wave interactions can be used to change the vortex charge of a beam.  
 For example in the weak pump depletion regime of  second harmonic generation the amplitude of the envelope of the harmonic field at frequency $\omega_2$ can be written as $A_2\sim A_1^2.$ Thus an input field with charge $m$ of the form $A_1\sim e^{\imath m\phi}$ generates an output field $A_2\sim e^{\imath 2 m\phi}$ with twice the charge.  This effect has been demonstrated experimentally 
by several groups using optical vortices created by diffraction from a hologram, and then allowing the beam to pass through a frequency doubling crystal\cite{so1993,pa1996,pi1997,pi1998,to1998b}.

In this work we describe a different approach to the generation of vortices in second harmonic generation that is based on frequency doubling of resonator modes with a vortical structure. Consider an empty resonator  with the pump beam mode matched to the lowest order transverse resonator mode (TEM$_{00}$ mode), which has a slowly varying amplitude at the cavity waist given by $u_{00} \sim\exp(-r^2/w_c^2),$ with ${\vec r}=x\hat x+y \hat y$ and $w_c$ the cavity waist. By changing the cavity tuning, and slightly tilting and displacing the pump beam,  we can couple to higher order transverse modes of the cavity. By appropriate alignment of the pump beam it is possible to couple to a single higher order transverse mode, such as $u_{10}\sim x\exp(-r^2/w_c^2)$ or $u_{01}\sim y\exp(-r^2/w_c^2)$  which have  edge dislocations. Following the cavity by an astigmatic mode-converter\cite{ref.tamm} the Hermite-Gauss modes can be efficiently converted into azimuthally symmetric Laguerre-Gauss modes with non-zero vortex charge, as was demonstrated by Snadden et al.\cite{ref.riis}.

As we show here it is also possible to generate a vortex mode directly, without using an astigmatic mode converter, by aligning the pump beam to give the desired superposition of $u_{10}$ and $u_{01}$ modes.
Let the pump beam be a displaced and tilted Gaussian. Ignoring unimportant constant amplitudes as well as any overall phase we have
$u_p=\exp{(-|\vec r - \vec r_p|^2/w_p^2)}
\exp{(i \vec q \cdot \vec r)}.$
Here $w_p$ is the pump beam waist, $\vec r_p=x_p\hat x + y_p\hat y$ is the transverse displacement of the pump beam, and $\vec q=q_x \hat x+q_y\hat y$ is the transverse wavevector that is proportional to the pump beam tilt  in the $x,y$ plane. 
The lowest order odd cavity mode that the pump couples to can be written as
\begin{equation}
u= o_{10} a(\nu-\nu_{10}) u_{10} + o_{01} a(\nu-\nu_{01}) u_{01}
\label{eq.ab}
\end{equation}
where $o_{mn}\sim\int dx dy~ u_p^* u_{mn}$ is an overlap integral, 
and $a(\nu-\nu_{mn})$ is a complex coefficient that depends on the difference between the pump frequency $\nu$ and the resonant frequency of the mode $\nu_{mn}.$ The amplitude and phase of the overlap coefficients can be independently varied by adjusting the pump beam\cite{ref.dza}. Performing the integrals we find $o_{10}=h (2 x_p/w_p - i k_x w_p)$ and
$o_{01}=h (2 y_p/w_p - i k_y w_p),$ where $h$ is an unimportant common factor. By adjusting the pump beam we can obtain $o_{01}=o_{10}\exp{(i\pi/2)}$ which results in vortex generation when  
$a(\nu-\nu_{10})=a(\nu-\nu_{01}).$

\begin{figure}[!t]
\begin{center}
\includegraphics[width=.6\textwidth]{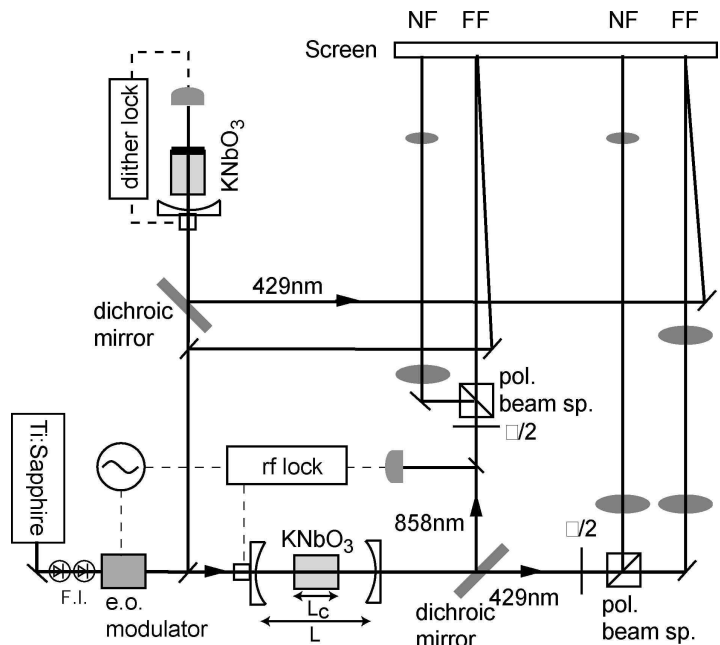}
\end{center}
\vspace{-.8cm}
\caption{Experimental setup. }
\label{fig.experiment}
\end{figure}
Vortices were generated in this way using the experimental setup  shown in Fig. \ref{fig.experiment}.
A Ti:Sapphire laser at 858 nm generates a continuous wave fundamental beam with a power of up to 300 mW incident on the freqency doubling cavity. The beam is mode matched to a linear cavity with two $R=25$ mm end mirrors that contains a 1 cm long a-cut KNbO$_3$ crystal with anti-reflection coated ends. The input and output mirrors had T$_{\rm 858nm}=4.8\%, 0.04\%$ and T$_{\rm 429 nm}=92\%$ so that only the fundamental field was resonant in the cavity.
 Phase matching was controlled by varying the temperature of the crystal. With the distance between the mirrors set to $L_{cf}=R+\frac{n_c-1}{n_c}L_c = 30.7$ mm   for confocal operation ($n_c$, $L_c$ are the crystal refractive index and length), and the crystal temperature tuned for large phase mismatch so no harmonic beam was generated, a cavity finesse of about 80 was measured. This agrees well with the theoretical value of $F=84$ that was calculated using measured values of 
the crystal losses. With the crystal temperature tuned for optimum phase matching  up to about 60  mW of 429 nm light was generated in a TEM$_{00}$ mode.

\begin{figure}[!t]
\begin{center}
\includegraphics[width=.6\textwidth]{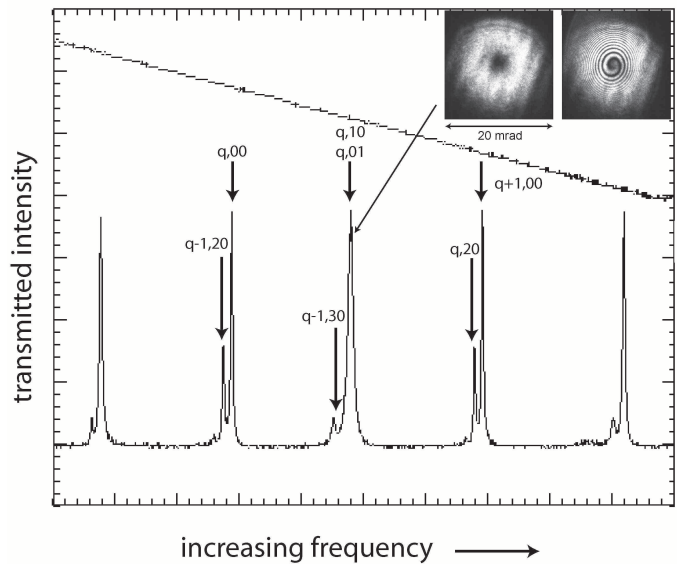}
\end{center}
\vspace{-.6cm}
\caption{Scan of resonator transmission with theoretical positions of the modes marked by arrows. $q$ denotes the axial mode index.  The upper line is the linear ramp voltage applied to the cavity piezo,and the inset shows the far-field structure of the fundamental field at the odd mode resonance.  }
\label{fig.modes}
\end{figure}

The cavity length was then reduced by 1.63 mm which resulted in the appearance of higher order transverse modes in the cavity transmission spectrum as seen in Fig. \ref{fig.modes}. Using the measured free spectral range as a scaling parameter the theoretically calculated frequencies of the first few higher order $(q,m,n)$ modes have been  indicated in the figure. The observed resonance frequencies agree to within  a few percent with the calculated values.  
The cavity was locked to a $(q,1,0)$ resonance using the rf sideband technique\cite{ref.pdh} which resulted in stable generation of a 
unit charge vortex mode as seen in the inset of Fig. \ref{fig.modes}.

\begin{figure}[!t]
\begin{center}
\includegraphics[width=.6\textwidth]{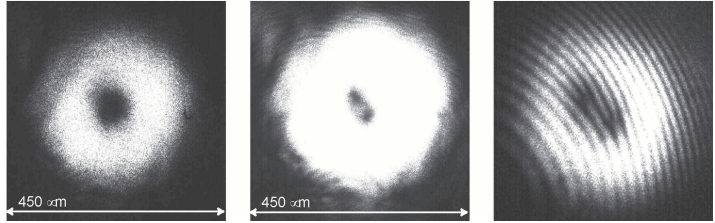}
\end{center}
\vspace{-.5cm}
\caption{Near field images of the harmonic charge 2 vortex. The central frame is overexposed to reveal the vortex splitting, as verified by the far-field interferogram on the right. }
\label{fig.charge2}
\end{figure}

\begin{figure}[!t]
\begin{center}
\includegraphics[width=.6\textwidth]{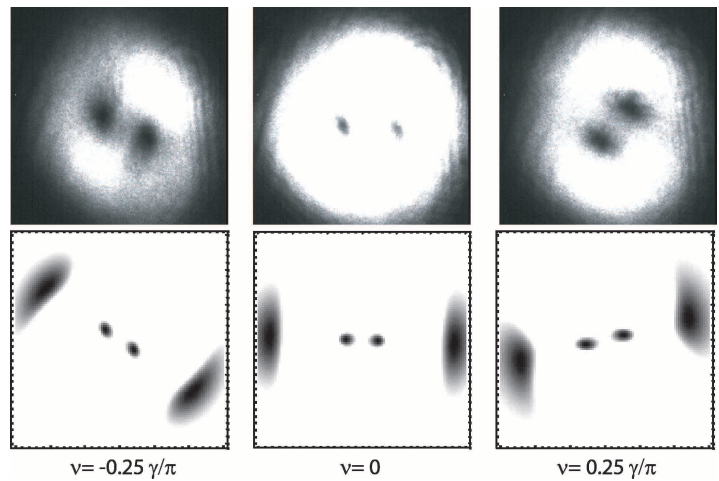}
\end{center}
\vspace{-.6cm}
\caption{Rotation of harmonic vortices observed in far-field (top row) and calculated from Eq. (\ref{eq.ab}) (bottom row). The columns are labeled 
with the pump beam frequency  which increases from left to right.}
\label{fig.rotation}
\end{figure}

The phase structure of the harmonic field was 
observed by interference with a TEM$_{00}$ beam generated in a second doubling cavity. 
As expected the second harmonic beam contained a doubly charged vortex
as seen in Fig. \ref{fig.charge2}.  
The detailed structure of the beam had a sensitive dependance on resonator alignment. Careful alignment of the crystal position resulted in observation of a  
doubly charged core region, although there was an apparent tendency for the vortices to repel each other so that a small 
vortex separation remained as seen in Fig. \ref{fig.charge2}.
We attribute the splitting to the topological instability of $m>1$ vortices\cite{mamaevprl}.
Adjustments to the crystal and/or pump beam alignment resulted in  the core splitting into two well separated singly charged vortices, with a controllable relative orientation and separation.
It was also found that when the resonator was aligned so that the cores were well separated, the relative orientation angle of the cores rotated in a repeatable fashion as the resonator was tuned across the $(q,1,0)$ resonance, as seen in Fig. \ref{fig.rotation}. 

The observations of vortex rotation  can be explained by taking account of the 
cavity astigmatism due to crystal birefringence.
Let the fundamental beam propagate along $z$ (a-axis of KNbO$_3$) and be polarized along $y$ (b-axis).
Beams propagating in the  $x-z$ plane correspond to ordinary polarized rays with an index $n_b=2.279$ at 858 nm and the temperature set for phase matching. We will label the $x-z$ plane modes with the first transverse index $m$ (the $x$ axis is horizontal in the figures).   
Beams propagating in the  $y-z$ plane correspond to extraordinary polarized rays with an index $n(\theta)=n_b(1+\tan^2\theta)^{1/2}/(1+(n_b/n_a)^2\tan^2\theta)^{1/2}$
where $\theta$ is the angle of the ray with respect to the $y$ axis 
and $n_a=2.238$ We will label the $y-z$ plane modes with the second transverse index $n.$  The effective crystal thickness for $x-z$ plane modes is $L_c/n_b$ while the 
effective  thickness for $y-z$ plane modes is $L_c/\tilde n,$ where
$\tilde n =  n_a^2/n_b$ is the effective index\cite{ref.siegman}.  
Since $\tilde n < n_b$ the crystal is effectively longer in the 
$y-z$ plane and these modes have lower resonance frequencies. A short calculation shows that the frequency splitting can be written as 
\begin{eqnarray}
\nu_{qm0}-\nu_{q0n}&=&\frac{c}{2\pi L_0}
\left\{ 
(m+1)\cos^{-1}\left(\frac{\delta}{R}\right)
\right.\nonumber\\
&&
\left.
-(n+1)\cos^{-1}
\left[\frac{\delta}{R}+\frac{L_c}{R}\left(\frac{1}{\tilde n}-\frac{1}{n_b} \right)
\right]
\right\}
\label{eq.split}
\end{eqnarray}
where $L_0=R+ [(n_c^2-1)/n_c]L_c +\delta$ with $\delta=L-L_{cf} $ the change in cavity length from confocality. 
For our exprimental parameters the shift given by Eq. (\ref{eq.split}) is about 15 MHz, which is a non-negligible fraction of the cavity resonance which has a FWHM of $2\gamma=2\pi \times 42~\rm  MHz.$ Indeed, close inspection of Fig. \ref{fig.modes} reveals that the odd order transverse modes are somewhat wider than the lowest order modes, in agreement with Eq. (\ref{eq.split}). The modal  coefficient $a$  that appears in Eq. (\ref{eq.ab}) can be written as  
\begin{equation}
a(\nu-\nu_{mn})=\left[1+(4 F^2/\pi^2)^2 \sin^2(\pi(\nu-\nu_{mn})/(2\gamma F)) \right]^{-1/2}  e^{\imath\tan^{-1}(\chi_{mn})},
\label{eq.amplitudes}
\end{equation}
where $\chi_{mn}= f \sin[\pi (\nu-\nu_{mn})/(F \gamma)]/\{1-f\cos[\pi (\nu-\nu_{mn})/(F \gamma)]\}$, and $f=1-\pi/F.$
For our experimental parameters the relative phase shift
when the resonator is tuned to the midpoint between the  $(10)$ and $(01)$ resonances is
 $\tan^{-1}\chi_{10} - \tan^{-1}\chi_{01}\sim  40^\circ.$

 The simulations of vortex rotation shown in Fig. \ref{fig.rotation} were obtained by
calculating $u$ from Eq. (\ref{eq.ab}), squaring the field, and  adding a small constant offset which resulted in splitting of the vortices. The phase difference between $o_{10}$ and $o_{01}$ was set to $\pi/2-40^\circ$  to give the correct $\pi/2$ phase shift for $\nu$ at the midpoint between the modes, after which the frequency $\nu$ was tuned across the resonance. 
We observe  qualitative agreement between the observed rotation, and the calculated results.  The reason for the calculated rotation being somewhat smaller than that observed may be that the calculations correspond to the near field while the observations shown in Fig. \ref{fig.rotation} were recorded in the far field. Additional observations revealed that the rotation angle was noticeably smaller in the near than in the far field. This is fully consistent with the propagation induced rotation of a vortex pair\cite{so1993}. It should be emphasized however that the observed dependence of the rotation angle on the cavity tuning is not a propagation effect, but rather due to the interference of 
nondegenerate cavity modes with amplitudes given by Eq. (\ref{eq.amplitudes}).
We also note that the shape of the vortex cores becomes elliptical on either side of the resonance. This may be due to the unequal changes in amplitude and phase of $a$ which implies that we are generating  and doubling  
anisotropic vortices\cite{torner_ani}. 

In conclusion we have observed creation, doubling, and splitting of vortices in a second harmonic generating resonator. Tuning of the resonator across a pair of non-degenerate transverse modes results in rotation of the harmonic vortex pattern.

This work was supported by National Science Foundation grant 0200372.
W. K.  acknowledges support from the U.S. Army Research Office.
M. S. is an A. P. Sloan Foundation fellow.

\end{document}